\documentclass[final,3p,times,twocolumn,authoryear]{elsarticle}
\usepackage{amssymb}
\usepackage{lipsum}
\journal{Computational Materials Science}
\begin{document}
\begin{frontmatter}
 \title{Assessment of approaches for dispersive forces employing graphone as a case study}%\tnoteZxref{label1}}
%%\author{author\corref{cor1}\fnref{label2}}
%% \ead{email}
%%\ead[url]{www}
 %%\cortext[cor1]{WWWW}
%\title{}
\author[label1]{Magdalena Birowska}
\author[label3]{Maciej Marchwiany}
\author[label4]{Claudia Draxl}
\author[label2]{Jacek A. Majewski}
 \address[label1,label2]{University of Warsaw, Faculty of Physics, 00-092 Warsaw, Pasteura 5, Poland}
  \address[label3]{Interdisciplinary Centre for Mathematical and Computational Modelling (ICM), University of Warsaw, Pawinskiego 5a, 02-106 Warsaw, Poland}
  \address[label4]{Theoretische Festk\"orperphysik, Humboldt-Universit\"at zu Berlin, Zum Gro{\ss}en Windkanal 6, 12489 Berlin, Germany}
%\ead[label1]{Magdalena.Birowska@fuw.edu.pl}
%% \address{Institute of Theoretical Physics, Faculty of Physics, University of Warsaw, Poland\fnref{label2}}
%%\ead[url]{www.fuw.edu.pl/~birowska}
%\ead[label2]{Jacek.Majewski@fuw.edu.pl}
%%\ead[url]{www.fuw.edu.pl/~birowska}
\begin{abstract}
We have studied two interchange layer systems, (i) free standing partly hydrogenated graphene (graphone), and (ii) graphone on the Nickel (111) surface, to assess various density functional theory based computational schemes incorporating van der Waals forces. The various van der Waals methods have been employed ranging from the semi-empirical force-field-like correction of Grimme, through non-local van der Waals density functionals, up to the functionals involving exact exchange and the random phase approximation for the correlation. Generally, all computational schemes lead to a similar qualitative picture of hydrogen layer physisorption and chemisorption to graphene. The largest discrepancies between the approaches emerge for the energetics of the investigated systems. Our studies shed light on the physical mechanisms of graphene hydrogenation both in vacuum and in the proximity of metallic surface. In particular, it is revealed that the adsorption of hydrogen atoms affects the nature of the bonding between graphene and the Ni(111) surface, from the weak to strong semi-covalent bonding. On the other hand, it turns out that the adsorption of hydrogen layer to graphene is stronger in the presence of the metallic surface. \end{abstract}
\end{frontmatter}
\section{Introduction}
Graphene, a precursor of two-dimensional materials, has attracted  great deal of attention in recent years, mostly due to remarkable properties  and whole plethora of potential applications [\cite{GeimNM}]. However, the absence of the band gap limits prospects of graphene based electronics. The change of graphene electronic properties can be induced through the suitable functionalization [\cite{Acta2011}], which turns out to be a promising route for the band gap engineering in many structures, \textit{e.g.}, Carbon-like systems [\cite{AIPMilowska,Acta2009}].
Specifically, the fully hydrogenated graphene with stoichiometry CH, \textit{i.e.}, with all carbon atoms saturated with hydrogens in such a way that two triangular sublattices have adsorbed hydrogens on the opposite sites of the graphene layer was successfully synthesized [\cite{Yang2016154}] and named \textit{graphane}. Graphane exhibits non-magnetic behavior with the band gap measured to be of the order of 4.5 eV [\cite{PhysRevB.75.153401}], in accordance with the earlier theoretical predictions[\cite{Elias610,Yang2016154}]. 

Especially interesting case of graphene's functionalization is incomplete hydrogenation, particularly one where only one of the graphene's triangular sublattices is hydrogenated on one side of the layer, resulting in 50 $\%$ of hydrogenation with stoichiometry C$_2$H. Such material, referred nowadays to as \textit{graphone} was first proposed theoretically by Zhou \textit{et al.} in 2009 [\cite{Zhou}]. Further, it has been predicted that graphone possesses indirect band gap of the order of 0.46 eV, and may exhibit also magnetic properties [\cite{Zhou, Zhang}]. The magnetic ordering as well as the bandgap can be controlled by the hydrogenation patterns. This makes hydrogenated graphene a promising material not only for electronic but also for future spintronics applications [\cite{Ray}]. Hydrogenation of graphene can also be important for other fields such as hydrogen storage [\cite{PhysRevB.77.035427}], Large scale production of an inexpensive graphene [\cite{10.1021}], or defect manipulation [\cite{PhysRevB.77.035427, 10.1021}].

Hydrogenated graphene can be synthesized by various experimental techniques [see the review \cite{C3CS60132C}]. One of the most common ways is to use graphene on a substrate (such as Au/Ni(111) [\cite{ADMA}], Ir(111), Pt(111) [\cite{10.1021/jp106361y}], SiC [\cite{10.1021/ja902714h}] and subsequent hydrogenation process. It has been recently demonstrated that graphone can be successfully synthesized on Ni(111) surface in the so-called reversible hydrogenation process [\cite{CHEM:CHEM201404938,  Bahn2017}], which turns out to be  crucial for hydrogen storage [\cite{10.1021}], as well as for controlling the electrical transport in graphene [\cite{Elias610}, \cite{Bostwick2009}]. Moreover, the partial one-sided hydrogenation of a graphene layer has recently been obtained by splitting of intercalated water 
in the Ni(111) supported graphene at room temperature [\cite{doi:10.1021/acsnano.6b00554}].

The crucial role of a substrate for the graphene hydrogenation process has been pointed out [\cite{Ray, CHEM:CHEM201404938,10.1021/jp106361y}]. Therefore, the theoretical investigation increasing understanding of the physical mechanism stabilising graphone on metallic substrates would definitely facilitate the synthesis of required materials based on hydrogenated graphene. The prerequisite for reliable studies of energetics of such layered systems is the correct description of the dispersive van der Waals (vdW) forces [\cite{APP2011.birowska,Birowska2019}, \cite{PhysRevB.84.201401}]. However, there are numerous ways of treatment of the dispersive vdW forces in the density functional theory based computational schemes, ranging from \textit{ad hoc} empirical corrections to the highly rigorous schemes with sophisticated density functionals involving exact-exchange and random-phase approximation for correlation. Following the trend of the high throughput calculations, here we investigate how different treatments of the VdW interactions 
influence the description of the hydrogenation, employing graphone as the case study.

In this paper, the van der Waals forces are treated by three different computational schemes within DFT as implemented in the VASP computer package: 
\begin{itemize}
\item \textit{Ad hoc} force-field corrections based on method of Grimme (DFT-D), where the dispersion term is added to the conventional Kohn-Sham DFT energy [\cite{Grimme}]. In this class, indicated here-after as DFT-D we consider several parametrizations: (DFT-D3) [\cite{DFT-D3}], (DFT-D3-BJ) [\cite{DFT-D3-BJ}], (DFT-TS) [\cite{PhysRevLett.102.073005}]. Generally, the acronym DFT-D denotes the method of Grimme throughout the paper. The DFT-D method is based on the parameterizations of entirely atomic quantities, renormalized by fitting to a set of molecules.

\item Non-local correlation functional (NLC) proposed by Dion \textit{et. al.}[\cite{PhysRevLett.92.246401}]. Within this method, the correlation term of the exchange-correlation energy is separated into local and non-local contributions, where in the latter the vdW interactions are included. We use several versions of this method: vdW-DF2 functional of Langreth and Lundqvist [\cite{PhysRevB.82.081101}], and others known as "Opt" functionals (OptPBE-vdW, OptB88-vdW, and OptB86b-vdW). In the latter, the exchange part of the functional has been optimized for the correlation part [\cite{opt}]. 

\item Random phase approximation (RPA) to the correlation energy, in the computational scheme  of Adiabatic-Connection Fluctuation-Dissipation Theorem (ACFDT) [\cite{PhysRevB.13.4274,PhysRevB.15.2884}]. The ground state energy of ACFDT-RPA (E$_{RPA}=E_{corr.}^{ACFDT-RPA}+E_{EXX}$) is the sum of the correlation energy E$_{corr.}^{ACFDT-RPA}$ and the exact exchange energy E$_{EXX}$ (also includes Hartree and kinetic energy terms, and Ewald energy of ions) evaluated non self-consistently using DFT orbitals. For details of implementation in the and assessment of the methods accuracy see the papers [\cite{PhysRevB.77.045136,PhysRevLett.103.056401,PhysRevB.81.115126}]. We denote this approach as RPA scheme throughout this paper. Note, that this nomenclature has been also used elsewhere [\cite{PhysRevB.84.201401,0953-8984-24-42-424218}].
\end{itemize}

We use graphone as a prototype system for the assessment of the above mentioned approaches with vdW forces taken into account on one hand, and for the investigation of the role of metallic substrate on the energetics of the hydrogenated graphene on the other hand. Therefore, to determine the influence of the substrate effect on the adsorption process of the examined structures, we carry out first the calculations for freestanding graphone and graphone formed on Ni(111) surface substrate. Second, we investigate the adsorption of small molecules like H, F, etc. on studied layered structures. This allows us to assess the role of the vdW interaction treatment  on the wide range of physical processes in layered materials. 
Hence, these investigations can be viewed as a benchmark for the functionalizations or intercalation of the other two-dimensional (2D) structures or vertically stacked 2D materials, forming new class of widely studied nowadays van der Waals heterostructures [\cite{geim2013van,Birowska2019}]. 

The paper is organized as follows. In section 2, we present computational details. The results are presented and discussed in section 3, where we deal with the energetics and structural changes of freestanding graphone and graphone on the Ni(111) substrate. Finally, the paper is concluded in section 4.  

\section{Computational details}

\begin{figure}
\centering
\includegraphics[width=0.45\textwidth]{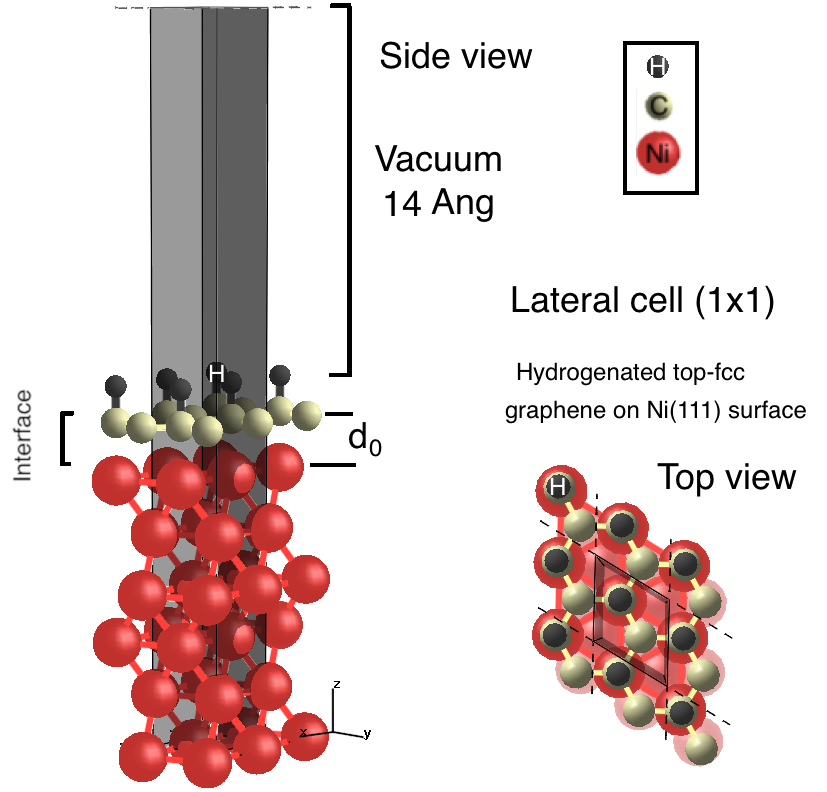}
\caption{\label{supercell} The schematic diagram of the supercell used in the calculations. The supercell consists of six monolayers of Ni and one monolayer of graphene. The hydrogen atoms adsorb on top of carbon atoms in graphene monolayer.}
\end{figure}

The calculations are performed within the spin polarized DFT [\cite{PhysRev.136.B864, PhysRev.140.A1133}] state-of-the-art computational scheme as implemented in the VASP package [\cite{PhysRevB.47.558, KRESSE199615}]. The electron-ion interaction is modelled by projector augmented wave pseudopotentials (PAWs) [\cite{PhysRevB.50.17953,PhysRevB.59.1758}] with GW potentials, which are believed to be important for RPA calculations, while the valence electrons are described with a plane wave basis set limited by a kinetic energy cutoff of 400 eV. The Brillouin Zone integration is done by using of $\Gamma$-centered $24\times24\times2$ Monkhorst-Pack sampling for all of the computational schemes employed except of the RPA. For the latter $19\times19\times1$ and $12\times12\times1$ Monkhorst-Pack samplings are used [\cite{PhysRevB.84.201401}] in the cases of graphone on Ni(111) substrate and freestanding graphone, respectively. The $(1\times1)$  lateral size of the supercell and 14 \AA{} of vacuum thickness is used in all of the calculations (see Fig. \ref{supercell}). The Ni(111) surface is modelled by a slab consisting of six monolayers (see Fig. \ref{supercell}). We consider here single-sided hydrogenation of graphene layer, where the hydrogen atoms fully occupy only one sublattice of graphene, as it is presented in Fig. \ref{supercell}B. The force convergence threshold is set to 0.002 eV/\AA. In the case of the RPA approach, the static calculations are performed for the atomic position optimization within the OPT-B88 vdW functional. The convergence criteria of RPA approach are adopted from Ref. [\cite{PhysRevB.84.201401} and see the references therein]. 

The structural optimization procedure has been done in two steps, first we optimize the lattice constant of fcc Ni structure for each of the method alone (the RPA lattice constant has been adopted from [\cite{PhysRevB.84.201401}]). Then, these Ni lattice constants suitable for corresponding method are used for the case of graphone on Ni(111) surface. Due to the fact that the graphene can be grown epitaxially on the Ni surface, the graphene has been stretched in our calculations to match the relaxed lattice constant of bulk Ni fcc. The calculations are carried out for the \textit{top-fcc structure} of graphene on Ni(111), which has been found to be the most stable structure [\cite{PhysRevB.84.201401,CHEM:CHEM201404938}]. Moreover, the positions of atoms of three bottom monolayers of Ni are fixed at the ideal bulk positions, whereas the other atomic positions of three upper monolayers of Ni and one monolayer of graphene at the top on the surface are fully relaxed for each of the hydrogen-graphene distance (only the z-coordinates of both hydrogen atom and carbon atom above H atom are fixed). In the case of the freestanding graphone, we optimize the lattice parameter of graphene for every computational scheme employed in this paper. Then, the suitable graphene's lattice constant of corresponding method is fixed for further calculations of the hydrogenation process.

\section{RESULTS}

Here, we present the main results of our studies. First, in Sec. \ref{energy}, we describe the energetic and structural issues of freestanding graphone. Then in Sec. \ref{graphoneSub}, we examine the energetic and structural properties of graphone on the Ni(111) surface. Finally in Sec. \ref{comparison}, we compare the results in order to determine the role of the substrate.

\subsection{Energetics and structural changes of freestanding graphone}
\label{energy}
\begin{figure}[h]
\includegraphics[width=0.5\textwidth]{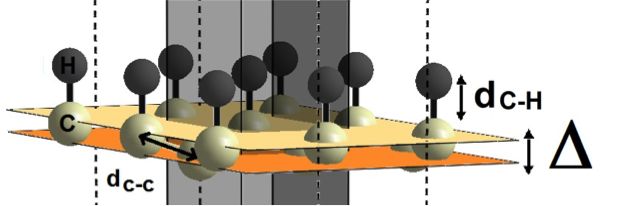}
\caption{\label{graphone} Freestanding single-side hydrogenated graphene (graphone). Only one sublattice of carbon atoms is saturated by hydrogen atoms 
(here indicated by the upper yellow plane, the second sublattice, hardly visible is within the orange lower plane). The dark grey balls, and the yellow balls indicate the hydrogen and the carbon atoms, respectively. 
$\Delta$ denotes the vertical distance between the plane containing carbon atoms from the sublattice A and the plane with carbon atoms belonging to the sublattice B. Therefore, $\Delta$ is the measure of the bowing in graphene layer after the adsorption of hydrogen atoms to one of the sub lattices.}
\end{figure}
\begin{figure}[b!]
 \includegraphics[width=0.5\textwidth]{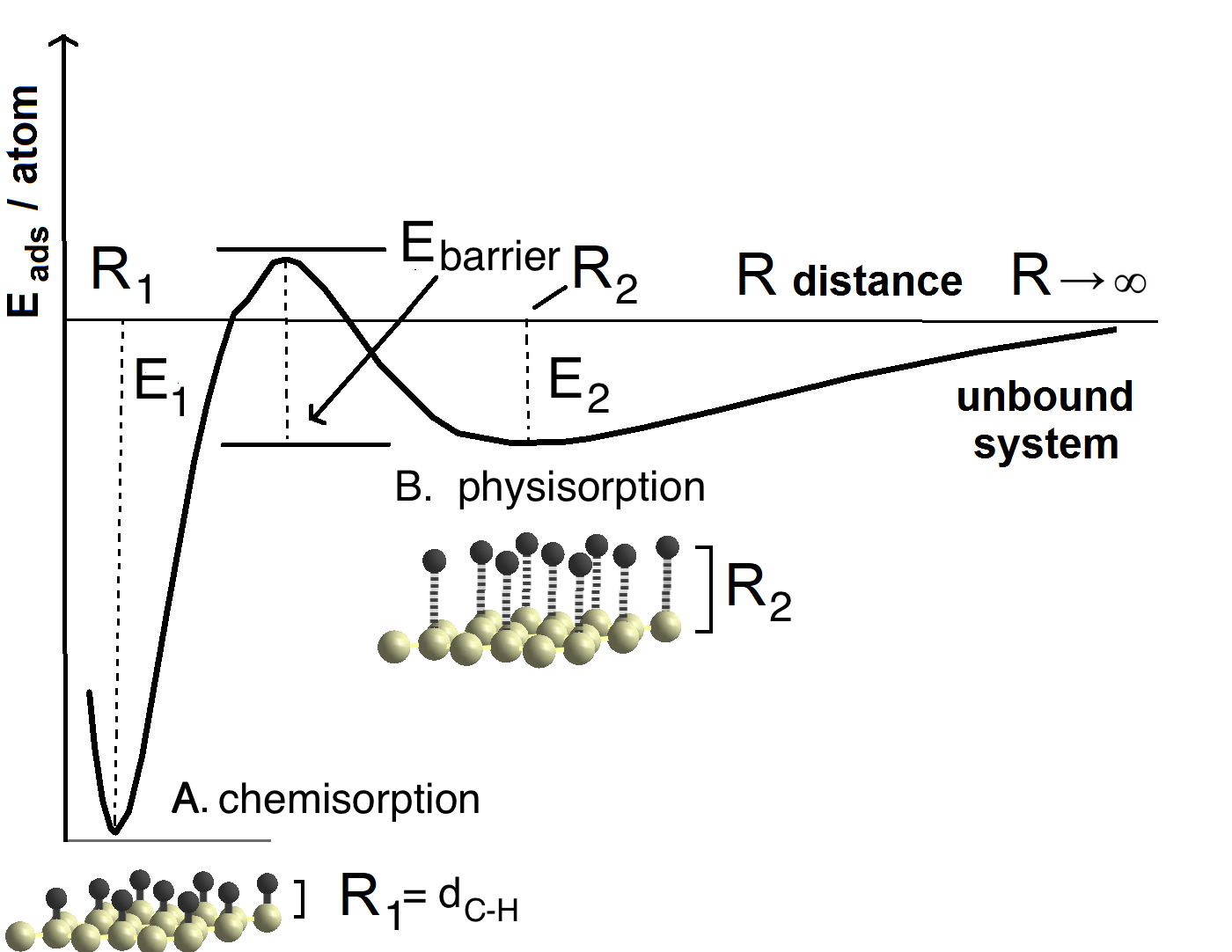}
 \caption{\label{bariera} Schematic diagram for the adsorption energy curve of hydrogen atoms on the graphene sheet versus the vertical distance $R$ between the hydrogen and graphene layer (\textit{i.e.}, the reaction coordinate). $E_1$, $E_2$, and $E_{barrier}$ denote the adsorption energy of chemisorption, physisorption, and energy barrier, respectively.  Two potential wells (two minima) are seen clearly. Deep narrow well around $R_1$ stands for the chemisorption of hydrogen atoms,
whereas wide shallow well in the neighborhood of $R_2$ corresponds to physisorption of hydrogen atoms. For each distance $R$ between the hydrogen and graphene,  all of the positions of atoms of the slab have been fully optimized. It is worth to mention that in the case of calculations where all of the coordinates of atomic positions are fixed, we obtain unphysical curve with physisorption well deeper than the chemisorption one.}
 \end{figure}

 \begin{figure}
\includegraphics[width=0.4\textwidth]{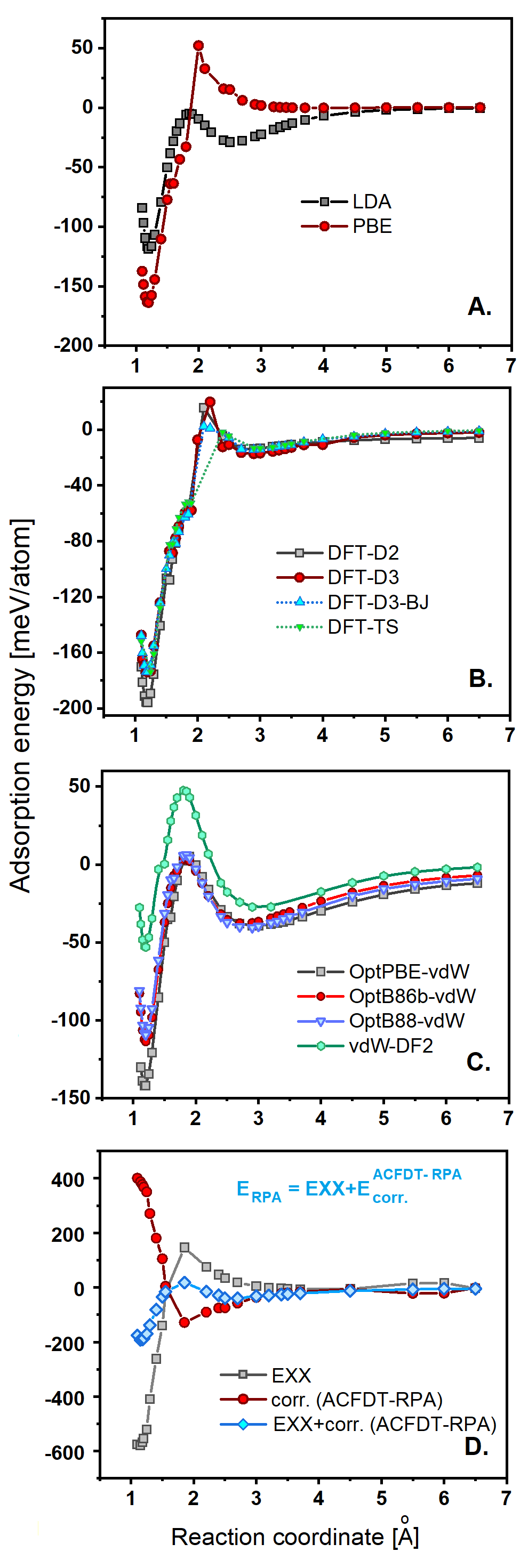}
\caption{\label{grH-vdW} Adsorption energy curves for the adsorption of the hydrogen atoms to the graphene layer as obtained with various computational schemes employed in this paper: (A) the semilocal functionals LDA and PBE, (B) force-field corrections (DFT-D), (C) non-local correlation functionals (NLC), and (D) the RPA energy curve (blue diamond data points) separated into exchange (black square data points) and correlation (red circle data points) contributions. The reaction coordinate $R$ is taken as the vertical distance, which has been fixed during the calculations, between the hydrogen atoms and the graphene layer. The lines are guide for the eye.}
\end{figure}  

\begin{figure*}[ht]
\centering
\includegraphics[width=0.80\textwidth]{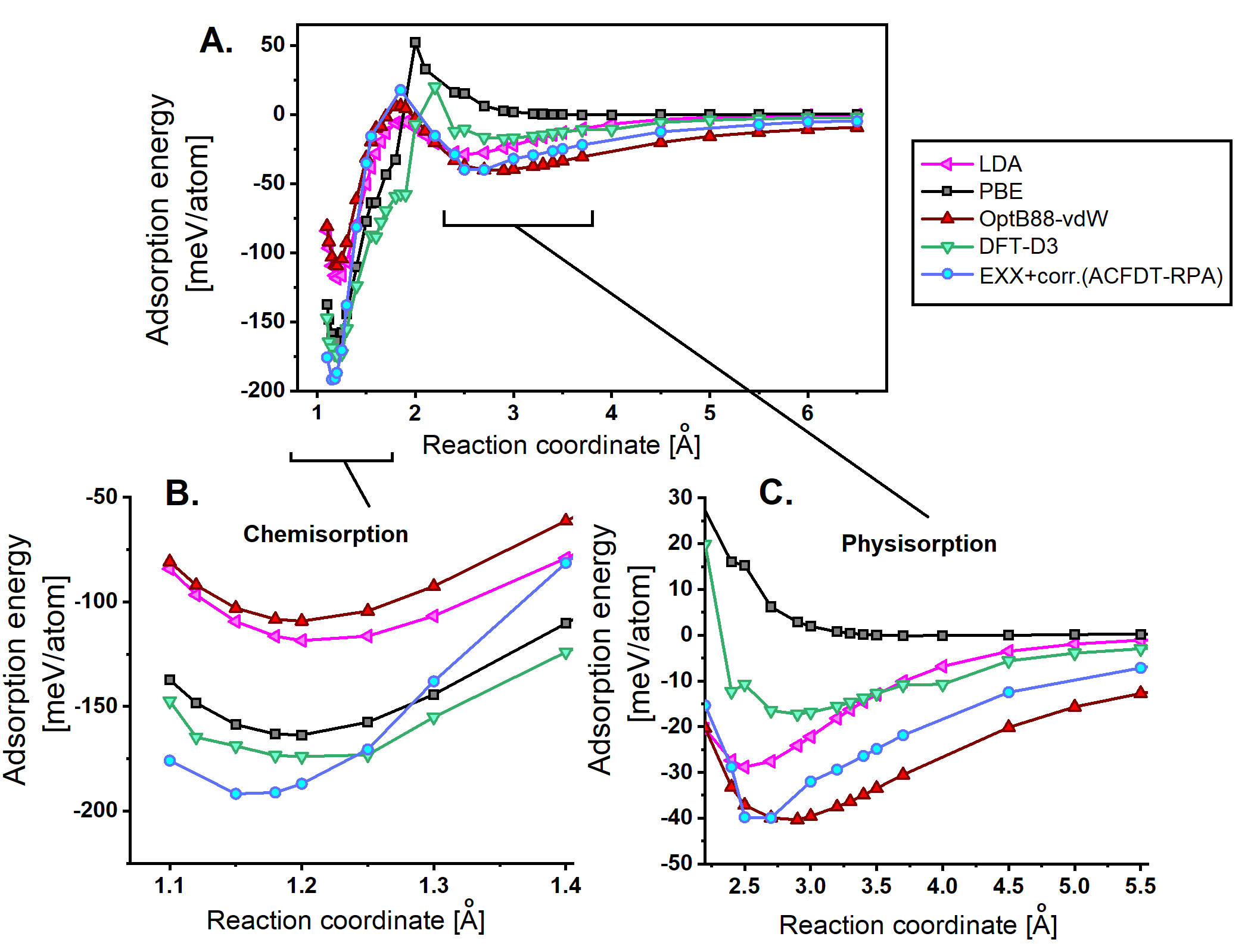}
\caption{\label{grH-all}(A) Comparison between various vdW approaches (RPA, DFT-D3, Opt-B88 functional) 
and standard semilocal functionals (LDA, PBE) for the adsorption energies of hydrogen atoms to the 
freestanding graphene for the whole range of calculated distances. (B) and (C): the zooms of the regions connected to chemisorption and physisorption, respectively. 
The reaction coordinate indicates the vertical distance $R$, between the hydrogen atoms and the graphene layer.}
\end{figure*}

The adsorption energy $E_{ads}(R)$ of the hydrogen atoms is calculated for a given distance $R$ between the hydrogen and carbon atoms that is considered as the reaction coordinate. For freestanding graphone $E_{ads}(R)$ is calculated as the difference between the total energy of graphone $E_{graphone}(R)$ (see Fig. \ref{graphone} and Fig. \ref{bariera}) and the total energy of the unbound graphene and hydrogen atoms $E_{graphene+H}^{unbound}$ (see Fig. \ref{bariera}) by using the formula:
\begin{equation}
	E_{ads}(R)=\frac{E_{graphone}(R)-E_{graphone+H}^{unbound}(R \mapsto \infty)}{N},
\end{equation}
where N is the total number of atoms in the supercell. By analogy, in the case of the hydrogenation of the graphene/Ni(111) system, the adsorption energy has the form:

\begin{equation}
E_{ads}(R)=\frac{E_{Ni-graphone}(R)-E_{Ni-graphene+H}^{unbound}(R \mapsto \infty)}{N}
\end{equation}
where, $E_{Ni-graphone}(R)$ denotes the total energy of the graphone on the Ni(111) surface with the distance between C and H atoms being equal to $R$, and $E_{Ni-graphene+H}^{unbound}$  indicates the unbound system of graphene/Ni(111) and hydrogen atoms. In order to obtain the energy for unbound system, we use the relation  $E(R)=E_0+A*e^{-R/B}$ for the reaction coordinate $R > $2.5 \AA{} (different for various approaches), and parameters $E_0$, $A$, $B$ fitted to the results of DFT calculations. Then, we assume that $E(R \mapsto \infty)=E_0$ is equal to the energy of unbound graphone (=$E_{graphone+H}^{unbound}$) and graphone on Ni(111) (=$E_{Ni-graphene+H}^{unbound}$). The negative values of the $E_{ads}$ indicate that system is bounded.

In the Fig. \ref{grH-vdW}, we present the hydrogen adsorption energy curves as the function of vertical distance between the hydrogen and carbon atoms from graphene layer, obtained with all computational schemes employed: (i) semilocal density functionals such as local density approximation (LDA) [\cite{PhysRevLett.45.566}] and the gradient-corrected Perdew-Burke-Ernzerhof (PBE) [\cite{PhysRevLett.77.3865}] (see \ref{grH-vdW}(A)), and (ii) approaches with vdW interactions taken into account such as DFT-D, OptPBE-vdW, OptB86b-vdW, OptB88-vdW, vdW-DF2, RPA. 

To assess the role of the non-local correlations, we plot their contribution to the $E_{ads}(R)$ in the Fig. \ref{grH-vdW}(D). The smallest chemisorption energy is obtained for RPA and DFT-D2 approaches and is equal to -192 meV/atom. Additionally the RPA approach gives the smallest distance between Hydrogen and Carbon atoms equal to 1.15 \AA{}. Our results reveal that inclusion of the non-local correlation lowers the chemical adsorption energy by 73 and 29 meV/atom in comparison to PBE and LDA functionals (see Fig. \ref{grH-all}(B)), respectively. It was previously shown, that the RPA covalent binding energies are at least as accurate as GGA functional for solids [\cite{PhysRevB.81.115126}], the latter is known to reproduce quite well the covalent binding. 

As can be seen in the Fig. \ref{grH-vdW}(B) and (C), within each vdW methods the shape of the energy landscape is qualitatively similar. The energy values for the covalent binding of hydrogen to graphene for the Grimme corrected methods (DFT-D) (see Fig. \ref{grH-vdW}(B))  are generally  smaller than for the semilocal functionals (see Fig. \ref{grH-vdW}(A) and the detailed comparison presented in Fig. \ref{grH-all}(B)). This is not surprising if one notes that the dispersion correction term is simply added to the conventional Kohn-Sham DFT energy. In the case of the various NLC functionals, the chemical adsorption energy varies in the range of -142 to -52 meV/atom. The energies are larger then obtained for the PBE functional.

The glance on the details of the $E_{ads} (R)$ curve in the physisorption range of reaction coordinates 
(see Fig. \ref{grH-all}(B)) reveals that the PBE functional provides no binding at the physisorption region ($R>$ 2 \AA{}), while the LDA predicts it at the distance 2.5 \AA{} with physisorption energy equal to 29 meV/atom. The existence of this minimum is an artifact of the LDA functional connected to the wrong decay of Kohn-Sham potential for $R \mapsto \infty$ [\cite{PhysRevB.84.201401}]. It is well known that the semilocal functionals have wrong exponential decay, whereas the proper van der Waals long-range behavior should be of polynomial type [\cite{PhysRevLett.90.066104, PhysRevLett.96.073201}]. All of the vdW approaches considered for freestanding graphone predict the physisorption energies at the distances lying in the range of 2.7 to 3.0 \AA{} (see Figs. \ref{grH-vdW} and \ref{grH-all}). Note that such distances are typical values for the van der Waals adsorbed graphene on transition metals [\cite{PhysRevLett.105.096801}]. The RPA approach and the NLC functionals give similar physisorption energy equal to -40 meV per atom (except for the vdW-DF2, where the physisorption energy is equal to -27 meV/atom), whereas in the case of the DFT-D approaches the energies are almost three times greater and of the order of -15 meV/atom, see Fig. \ref{grH-all}(C)).

Moreover, all of the applied methods of the treatment of vdW forces predict the energy barriers between the chemisorption and physisorption (covalent and non-covalent binding). The NLC and RPA approaches predict much greater barrier energies (40-75 meV per atom and 58 meV per atom, respectively) than DFT-D methods (15-40 meV per atom) (see Figs. \ref{grH-vdW} and \ref{grH-all}). Note, that the energy barriers are just approximate values especially for the RPA and DFT-D cases, for which it is rather hard to reach generally required convergence level for $R$ around 2.00 \AA.

\begin{table}
\small
\centering
\caption{Bonding energy $E_1$, bond length $R_1$, and the buckling of the graphene sheet after hydrogen adsorption $\Delta$ for the chemisorption of hydrogen atoms to the freestanding graphene for all computational schemes employed in the paper. Here, we present the results for fully optimized positions of all atoms. Thus, the values of $E_1$ may differ slightly from the values presented in Figs. \ref{grH-vdW} and \ref{grH-all}, where the values of $E_1$ are extrapolated from $E_{ads}(R)$ curve.}
\bigskip
\label{tab:grHtable}
\begin{tabular}{ | c | c | c | c  |}
\hline
\centering
Methods & $E_1$ [meV/atom]  & $R_1$ [\AA] & $\Delta$ [\AA] \\ 
\hline
\hline
 PBE & -224  & 1.17  & 0.33 \\
  LDA & -116 & 1.20     & 0.30  \\
  \hline
  DFT-D2 & -257  & 1.17    & 0.26 \\
DFT-D3 &  -174 & 1.19      & 0.26 \\
   DFT-D3-BJ &  -174 & 1.19     & 0.26 \\
   DFT-TS & -179  &  1.19  & 0.26 \\
   \hline
   OptPBE-vdW& -176  & 1.17  & 0.34\\
   OptB88-vdW&  -160 & 1.17  & 0.33 \\
   OptB86b-vdW & -155 & 1.18    & 0.33 \\ 
   vdW-DF2 & -122  & 1.16 & 0.34 \\
   \hline
\end{tabular}
\end{table}

Our results confirm that the non-local correlation functionals might be seen as promising candidates to reproduce qualitatively the same results for physisorption as more demanding RPA approach (see results for Opt-B88 and the RPA in Fig. \ref{grH-all}(C))\footnote{The computational load of the RPA approach is widely reported in literature \cite{PhysRevB.87.075111} and for the EXX part \cite{APP2016.Birowska}}, as it was previously mentioned in the paper [\cite{PhysRevB.84.201401}], for the case of the graphene on the Ni(111) surface. In addition, NLC functionals follow the trend of the long range behavior of the RPA approach (see Fig. \ref{grH-all}(C), and correctly reproduce the long range behavior of the vdW interaction [\cite{0953-8984-24-7-073201}], which is known 
from the fundamental theoretical considerations.

We have also carried out the calculations of the covalent bonding energies and corresponding structural 
parameters employing all considered in this paper computational schemes. In these calculations the full relaxation of atomic positions has been performed, \textit{i.e.}, carbon-hydrogen distance $R$, considered in the calculation of the $E_{ads}(R)$ as fixed reaction coordinate, has been also relaxed. The results are collected in the Table \ref{tab:grHtable}. Note that the chemisorption energies $E_1$ calculated from fully relaxed  geometry might slightly differ from those which  have been  extracted from $E_{ads}(R)$ curves and presented in Figs. \ref{grH-vdW} and \ref{grH-all}. Our results reveal that the chemical adsorption of the hydrogen atoms causes the buckling of the graphene layer (the distance $\Delta$ between two carbon sublattices constituting the graphene see Fig. \ref{graphone}). The NLC functionals give similar values as PBE functional, whereas DFT-D methods give by 0.07 \AA{} smaller values. In addition, the buckling effect appears only for covalent bonding, no buckling is observed for the physisorption region. 

To the best of our knowledge, there are no structural or energetic experimental data available for graphone system, thus, we can only compare our findings with the theoretical predictions. The authors [\cite{Zhou}] using the PBE functional, predicted the length of the carbon-carbon bond of graphone to be 1.50 \AA{} and carbon-hydrogen bond equal to 1.18 \AA{}, which are close to our findings for PBE functional giving 1.50 \AA{} and 1.17 \AA{}, respectively.

\subsection{Energetics and morphology of graphone on Ni(111) surface.}

\label{graphoneSub}
\begin{figure}[h]
 \centering
 \includegraphics[width=0.45\textwidth]{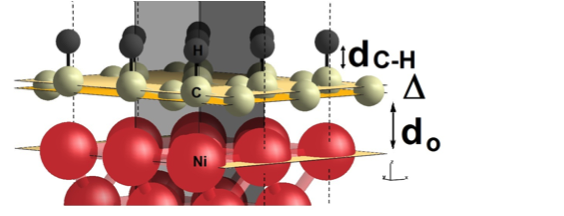}
 \caption{\label{barieraNi} Graphone on the Ni(111) surface. The equilibrium parameters $d_{C-H}=R_1$, $\Delta$ and $d_0$ denote the vertical distance between the hydrogen and carbon atoms, buckling parameter and the vertical distance between the graphene and Ni(111) surface, respectively.}
\end{figure}

\begin{figure}
 \centering
 \includegraphics[width=0.35\textwidth]{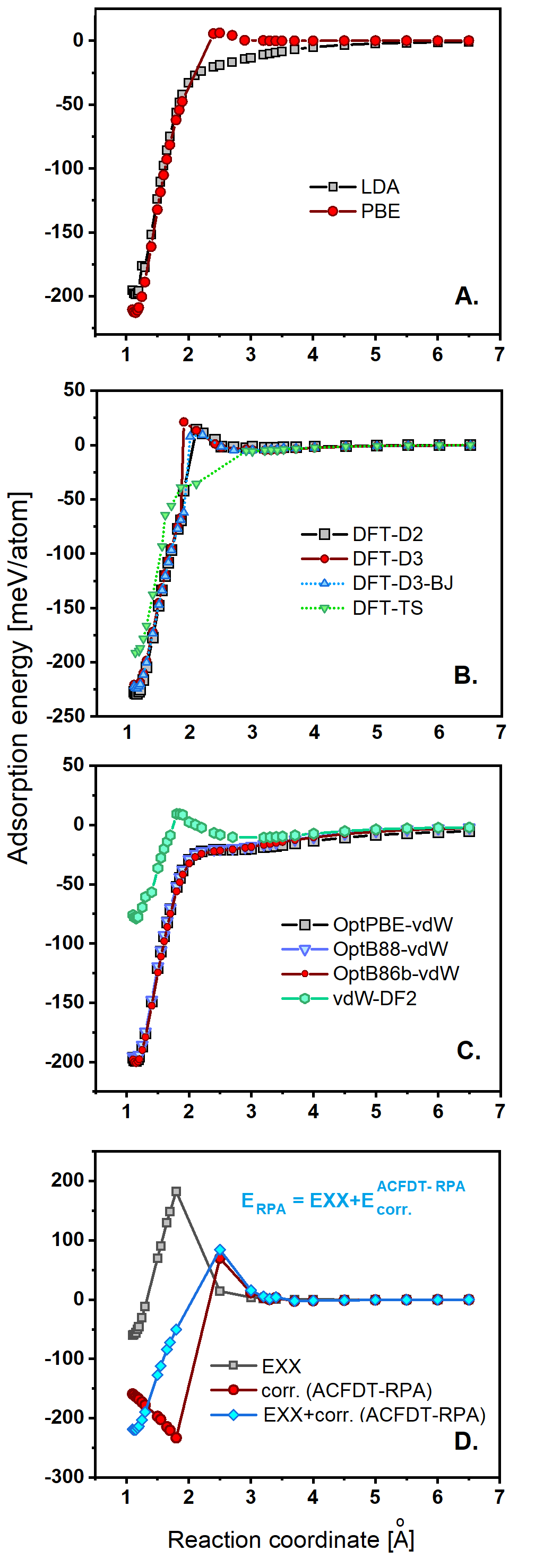}
 \caption{\label{NigrH}  Adsorption energy curves for the adsorption of the hydrogen atoms to graphene/Ni(111) surface as obtained with various computational schemes employed in this paper: (A) for the semilocal functionals  LDA and PBE, and the different vdW approaches: (B) force-field corrections (DFT-D), (C) non-local correlation functionals (NLC), and (D) the RPA energy curve (blue diamond data points) separated into exchange (black square data points) and correlation (red circle data points) contributions. The reaction coordinate $R$ indicates the vertical distance, which was fixed during the calculations, between the hydrogen atoms and the graphene layer. The lines are guide for the eye.}
\end{figure}
\begin{figure*}[]
 \centering
 \includegraphics[width=0.7\textwidth]{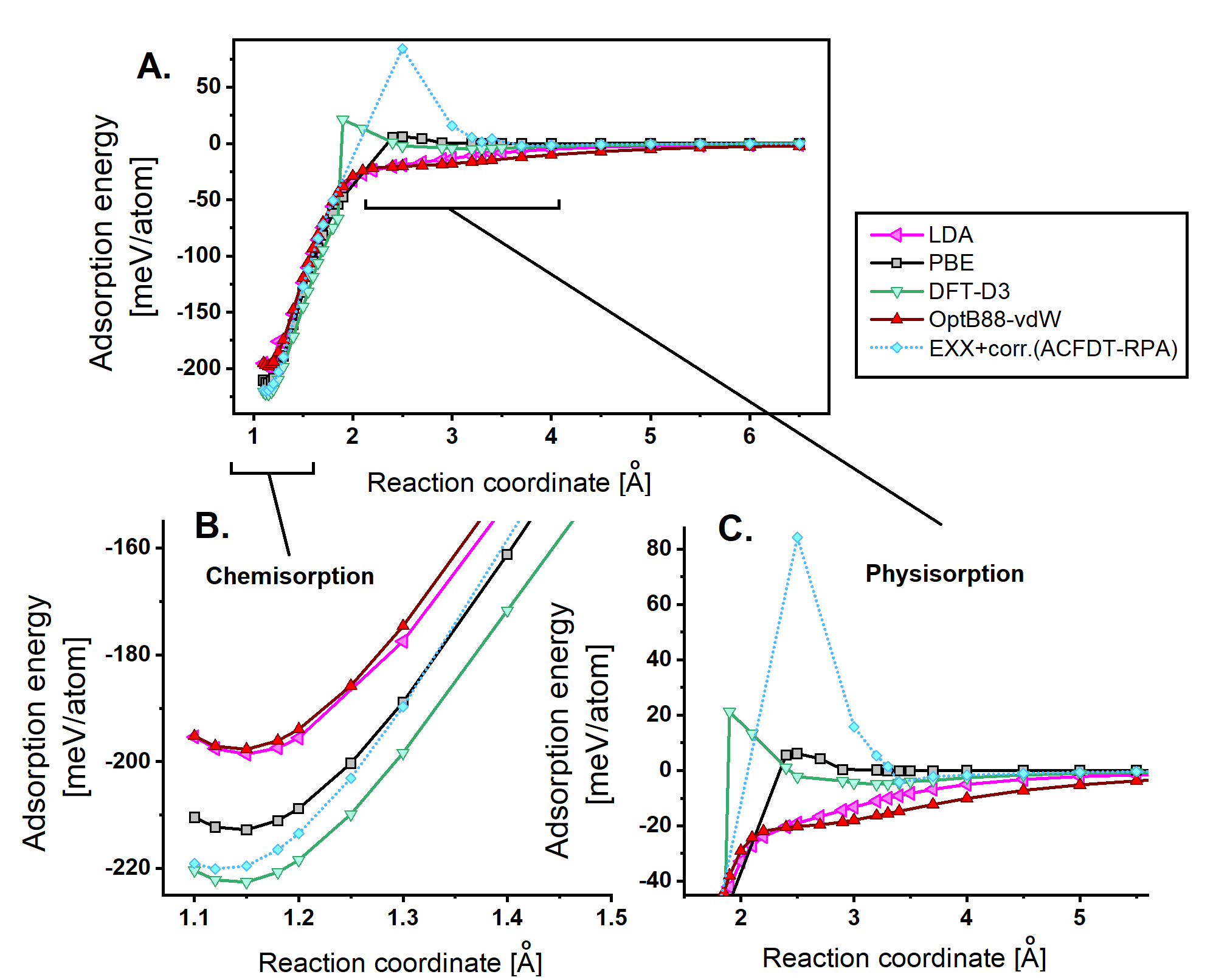}
 \caption{\label{NigrHall} Adsorption energy curves for the hydrogenation of the graphene/Ni(111) surface 
 for some of the vdW approaches employed in this paper. The magnifications of the energy curves  around chemisorption and physisorption regions are depicted in panels (B) and (C), respectively. Lines are guide for the eye. }
\end{figure*}
The hydrogen adsorption energy landscape on Ni(111) surface supported graphene, clearly depicts one potential well at the distance of 1.15 \AA{} between the hydrogen and carbon atoms, independently of employed computational scheme (see Figs. \ref{NigrH} and  \ref{NigrHall}(B)). Moreover, for the NLC functionals, we obtain the greatest chemisorption energies equal about -200 meV/atom (see Fig \ref{NigrH}(C)). The striking difference in adsorption energy curve occurs for vdW-DF2 in comparison to all other NLC functionals (see Fig. \ref{NigrHall} (C) and (D)).
Namely, the chemical adsorption energy is about 2.5 times greater here (with the energy equal to -78 meV/atom) than obtained with other methods. Previously, it has been also shown, that the vdW-DF2 are too repulsive at short distances [\cite{PhysRevLett.92.246401, PhysRevB.84.201401}]. The smallest chemisorpted energy equal to -229 meV/atom is obtained for DFT-D2 approach, which is just few meV/atom smaller than for the RPA approach being is -220 meV/atom (see Figs. \ref{NigrH} and  \ref{NigrHall}(B)). The RPA chemical adsorption energy has lower value than  energies obtained from the semilocal functionals (similarly as in the case of the freestanding graphone), as well as in the DFT-D approaches.

Note that DFT-D schemes and vdW-DF2 predict 
physical adsorption wells with very shallow minima in the energy range of -6 to -3  meV/atom  and -11 meV/atom, respectively, at the distances in the range of 2.9-3.2 \AA{}. In addition, the energy barriers are predicted only to occur in the case of the DFT-D, vdW-DF2 and RPA approaches, with the energies equal to 15-26 meV/atom, 20 meV/atom and 89 meV/atom,  respectively (see Fig. \ref{NigrHall}(C)).
\begin{table}
\small
\centering
\caption{Energy values and structural parameters for the chemisorption of hydrogen atoms on graphene/Ni(111) surface, for the case of fully relaxed atoms. Thus, the chemisorpted energies $E_1$ may differ from the values presented in Figs. \ref{NigrH} and \ref{NigrHall} (the vertical distance between the hydrogen and carbon atoms is fixed). All of the first row symbols are explained in the Fig. \ref{barieraNi}. The $d_0$ and  $\Delta$ values indicate the vertical distance between the graphene and Ni(111) surface, and the buckling of the chemical adsorption, respectively. $R_1$ is Carbon-Hydrogen bond length. All of the structural parameters are given in $\AA$.} 
\bigskip
\label{tab:tabNigrH}
 \begin{tabular}{ | c | c | c | c | c |}
\hline
\centering
Methods & $E_1$ [meV/atom] & $R_1$  & $d_0$  & $\Delta$ \\ \hline
\hline
 PBE & -213  & 1.15 & 1.75& 0.47   \\
 LDA & -188  & 1.15  & 1.70 &0.47 \\
\hline
 DFT-D2 & -216  & 1.14 & 1.73& 0.47 \\
 DFT-D3 & -223  & 1.14 & 1.73 & 0.47 \\
 DFT-D3-BJ&  -211 & 1.14 &  1.74  & 0.47 \\
 DFT-TS & -226  & 1.14&  1.74& 0.48 \\
\hline
 OptPBE-vdW & -176  & 1.14 &1.78 & 0.47 \\
 OptB88-vdW&  -172 & 1.14 & 1.77& 0.47 \\
 OptB86b-vdW& -188  & 1.15 & 1.77 & 0.47 \\ 
 vdW-DF2 & -158  & 1.14 &  1.88& 0.45 \\
%\hline
 %RPA=corr.+EXX &  & 1.12  & - & - \\
\hline
\end{tabular}
\end{table}

Similarly as in the case of the freestanding graphone, we have also carried out the full relaxation of the atoms, without any constraint of vertical distance between the hydrogen and carbon atoms. These results are collected in the Table \ref{tab:tabNigrH}. One can notice that in general the buckling of the carbon sublattices of graphene after hydrogen adsorption is equal to 0.47 \AA{}, independently of the method used. Note that the distance between the graphene sheet and the Ni(111) surface $d_0$ after Hydrogen adsorption is approximately about 1.8 \AA{}, which is the smaller value in comparison to the reported distance 2.2 \AA{} between the graphene and Ni(111) surface without hydrogen adsorption [\cite{PhysRevB.84.201401}]. This result reveals that hydrogen adsorption to the graphene/Ni(111) surface affects the nature of the bonding between the graphene and the Ni(111) surface. Specifically, it changes the type of the bonding between the carbon and the Ni atoms, from mainly weak van der Waals type of binding strong semi-covalent bonds.

\subsection{The influence of the Nickel substrate on the graphone}
\label{comparison}

\begin{table}
\small
\centering
\caption{The adsorption energies of hydrogen to graphene layer and structural parameters of freestanding graphone (gr-H) and formed on Ni(111) surface (Ni-gr-H). $\Delta E$ (in meV/atom) indicates the difference between the hydrogen adsorption energy to graphene on Ni(111) surface and freestanding one, whereas $\Delta d_{C-C}= d^{Ni-gr-H}_{C-C}-d^{gr-H}_{C-C}$, and $\Delta d_{C-H}= d^{Ni-gr-H}_{C-H}-d^{gr-H}_{C-H}$ are changes in the carbon-carbon and carbon-hydrogen bond lengths, respectively. $\frac{\Delta Ni−gr−H}{\Delta gr-H}$ indicates the ratio of buckling parameters for the two cases, and $\Delta d_{0}=d_{0}^{Ni-gr-H}-d_{0}^{Ni-gr}$ is the difference in the distance between the Ni and C layers in graphone on Ni, and graphene on Ni. Note, that the large value of $\Delta d_0$ for the vdW-DF2 stems from the large equilibrium distance between the graphene and Ni(111) surface equal to 3.7 \AA{}. Similar value for this distance has been reported previously (\cite{PhysRevB.84.201401}] and references therein). 
This originates mostly from too repulsive character of the vdW-DF2 functional. All the structural values are given in $\AA{}$.} 
\bigskip
\label{tab:NigrHdist1}
  \begin{tabular}{ | c | c | c | c | c | c |}
\hline
\centering
Methods & $\Delta E$& $\Delta d_{C-H}$ &  $\frac{\Delta_{Ni-gr-H}}{\Delta_{gr-H}}$  & $\Delta d_{C-C}$  & $\Delta d_{0}$ \\ 
\hline
\hline
 PBE & 10&-0.03 & 1.4  & 0.06 & -0.40  \\
 LDA & -71&-0.05  & 1.6  & 0.04  & -0.37  \\
  \hline
 DFT-D2 & 41& -0.03  & 1.4  & 0.03  & -0.42  \\
 DFT-D3 & -49 &  -0.06 &  1.8  &0.06 & -0.49 \\
 DFT-D3-BJ &-36& -0.05  &  1.8  & 0.05 & -0.45  \\
 DFT-TS & -46 &-0.06     &  1.8  &  0.04   & -0.51\\
  \hline
 OptPBE-vdW & 2& -0.02  & 1.4  & 0.06  & -0.46 \\
 OptB88-vdW & -14& -0.03   &  1.4  & 0.05  &  -0.45  \\
 OptB86b-vdW& -34& -0.03 & 1.4   & 0.04  & -0.41  \\ 
 vdW-DF2 &-36 &  0.03 &   1.3&  0.08& -1.82  \\
\hline
  \end{tabular}
\end{table}

In this section, we present comparison of the energetics and structural properties of the free standing graphone (indicated as ‘gr-H’) and the graphone formed on the Ni(111) surface (indicated as ‘Ni-gr-H’) as calculated within all employed in this paper computational schemes (see Table \ref{tab:NigrHdist1}). First, we focus on the adsorption energy of the hydrogens to graphene layer. The presence of the Ni(111) substrate changes this energy by the magnitude $\Delta E$ that is depicted in the second column of the Table \ref{tab:NigrHdist1}. As one can see, $\Delta E$  is negative in all but two cases (PBE and OptPBE-vdW), just indicating that the presence of Ni favours energetically the adsorption of the hydrogen. In addition, the non-local correlation functionals do not predict the energy barriers for the chemical adsorption of the hydrogen atoms on the graphene Ni(111) surface, whereas the force-filled corrections based on Grimme method (DFT-D) predict this barriers to be very small. Generally, this indicates that the formation of graphone is more energetically favourable on the Ni(111) surface than in vacuum.

The structural properties of the free standing graphone and the graphone formed on the Ni (111) surface are exemplified in Table \ref{tab:NigrHdist1} by the geometrical parameters such as carbon-carbon in-plane bond length $d_{C−C}$, and carbon-hydrogen vertical bond length $d_{C−H}$. The differences in $\Delta d_{C-C}=d_{C-C}^{Ni-gr-H}-d_{C-C}^{gr-H}$, and $\Delta d_{C-H}=d_{C-H}^{Ni-gr-H}-d_{C-H}^{gr-H}$ are depicted in the column 5 and 3 of the Table \ref{tab:NigrHdist1}, respectively. It is clearly seen that for graphone on Ni(111) the $C-C$ bonds are elongated and $C-H$ bonds shortened, independently on the computational scheme employed. The shortening of the $C-H$ bond for system on Ni(111) surface is in accord with the lowering of the hydrogen adsorption energy described above. Moreover, one observes increase of the buckling parameter $\Delta$ by a factor of order of 1.8 in the graphone on Ni(111) in comparison to the freestanding one (see column 4 of Table \ref{tab:NigrHdist1}). It is also interesting to compare the distance between Ni(111) substrate and graphene and
graphone. As it is depicted in the last column of Table \ref{tab:NigrHdist1} the distance between carbon layer and nickel surface $d_0$ is considerably lower in the case of hydrogenated graphene (\textit{i.e.}, graphone). All this
reveals that the bonding mechanism between Ni(111) surface and carbon layer changes as the result of graphene hydrogenation and formation of graphone.

\section{Conclusions}

In summary, we have investigated the energetic and structural properties of the freestanding and formed on the Ni(111) surface graphone, by using various approaches incorporating vdW forces into DFT based total energy functionals. The investigated methods include the semilocal functionals (LDA and PBE), the semi-empirical 
force field-like Grimme corrections (DFT-D), non-local van der Waals density functionals (NLC), and the functionals employing the exact exchange and the random-phase approximation for correlation (RPA).

Closer examination of the hydrogen adsorption energy curves on freestanding graphene layer reveals two potential wells, referred to chemical and physical adsorptions, and energy barriers between physisorption and chemisorption energy minima for the LDA and all of the considered in this paper vdW methods. On contrary, the hydrogen adsorption energy curves on the graphene/Ni(111)
surface exhibit, generally, only one deep chemisorption potential well with no energy barriers for the chemical bond formation. In addition, the investigated approaches predict typically smaller (more negative) hydrogen chemisorption energies to the graphene/Ni(111) system in comparison to the freestanding graphene.  It strongly indicates that it would be easier synthesise graphone on nickel substrate than in the freestanding form.  Moreover, the adsorption of hydrogen to the graphene on the Ni(111) surface causes the decrease of the distance between the graphene and Ni(111) surface, and thus, changing the nature of the chemical bonding between the carbon and nickel atoms, from weak to strong semi-covalent one.

In addition, in the case of the covalent bonding, the RPA and DFT-D2 predict the lower chemisorpted energies than PBE functional (which is commonly considered to account fairly well for the covalent binding), 
whereas the NLC functionals predict the greater values of these energies than the PBE. In the case of the freestanding graphone, the similar energy landscapes in the physisorption region are obtained within RPA and NLC functionals.

\section*{Acknowledgments}

This work was supported by National Science Center (NCN) through the grant HARMONIA DEC-2013/10/M/ST3/00793. M. B. acknowledges the support of National Science Center, Poland through the grant SONATA UMO-2016/23/D/ST3/03446.
We have used computing facilities of PL-Grid Polish Infrastructure for Supporting Computational Science in the European Research Space, and acknowledge the access to the computing facilities of the Interdisciplinary Centre of Modeling (ICM), University of Warsaw. 

 %%\bibliographystyle{elsarticle-harv} 
% \section*{References}
\bibliographystyle{elsarticle-harv}
 \bibliography{graphone}
%\begin{thebibliography}{bibliografia}
%\end{thebibliography}
\end{document}